\documentclass[12pt]{article}

\usepackage{scicite}
\usepackage{graphicx}

\usepackage{times}

\newenvironment{sciabstract}{%
\begin{quote} \bf}
{\end{quote}}

\title{Technical Comment on ``The dark matter interpretation of the 3.5-keV line is inconsistent with blank-sky observations"}

\author
{Kevork N.\ Abazajian\\
\\
\normalsize{Center for Cosmology, Department of Physics and Astronomy,}\\
\normalsize{University of California, Irvine, CA 92697, USA}\\\\
\\
\normalsize{E-mail:  kevork@uci.edu.}
}

\date{}

\begin{document} 


\baselineskip24pt


\maketitle


\begin{sciabstract}
  I show that model dependencies in the analysis by Dessert, Rodd \& Safdi (2020) relax their claimed constraint by a factor of $\sim$20. After including conservative model choices, the derived limits are comparable to or slightly better than limits from previous searches. Further model tests and expansion of the data energy may enhance or relax sensitivity of the methodology.
\end{sciabstract}

\section*{Introduction}
Since 2014, an unidentified X-ray line (UXL) that is attributable to dark matter (DM) decay has been detected in a large number of observations (for a review, see, e.g. \cite{Abazajian:2017tcc}). The paper by Dessert, Rodd \& Safdi (2020) \cite{Dessert:2018qih} (hereafter DRS) uses a large sample of observations by the \emph{XMM-Newton} X-ray Space Telescope to search for and place constraints on the presence of this line while masking the targets of those observations, typically compact stellar remnants and other X-ray luminous astrophysical sources. The paper by  DRS concludes with the claim, ``Our analysis rules out the decaying DM interpretation of the previously observed 3.5-keV UXL because our results exclude the required decay rate by more than an order of magnitude." Given limitations in model selection and data selection in that analysis, I show that the claimed limit of DRS is over-stated by at least a factor of 20. 

\section*{Model Dependencies}
When considering the models that contribute to the interpretation of observations that place a limit on model features, and when there exists no prior information that differentiates or prefers one of those models, the model with the most conservative limit should be selected. Placing constraints on the DM component of our Galaxy depends on its distribution, which is provided by models constrained by other data as well as physical constraints and expectations from simulations of our Galaxy's formation. In addition, model choices of the continuum and line X-ray emission in the regions of interest are also important to interpret new X-ray signals that may exist in the data. 

\subsection*{Dark matter density normalization and profile}
Constraints on flux from DM is highly dependent on the amount of DM inferred to be present in the observations. DRS choose a local density of the DM of $0.4~\rm GeV/cm^3$ in our Milky Way Galaxy's halo. The inferred density has a wide range of values from different techniques. Among the most robust determinations of the local dark matter density from the velocities of K-dwarf stars measured by the Sloan Digital Sky Survey, which finds $0.28\pm0.08~\rm GeV/cm^3$\cite{Zhang:2012rsb}. Just as importantly, hydrodynamic simulations of MW-like galaxies also show the presence of a core in the dark matter density profile with a size of roughly a kpc \cite{Kuhlen:2012qw} in response to the presence of the stellar bar \cite{Weinberg:2001gm}. A similar core also arises in simulations with a fixed disk and bulge potential \cite{Robles:2019mfq}. The effect of a cored  profile was included in a test of in the SM of DRS, but not the model dependence of the local density. Conservatively adopting the lower local density above and the cored density profile combine to relax the line limit in DRS by a factor of approximately three.

\subsection*{Other X-ray Lines within the Energy Range}
DRS chose to analyze a very narrow energy range, from 3.3 to 3.8 keV, even though the energy resolution of the instruments is $\sim$0.1 keV. Even within that narrow range, two groups of non-DM lines are known to contribute: at 3.3 keV and 3.7 keV. Many of the lines of sight point toward the Galactic plane and by-design include astrophysical sources that could contribute high-energy line transitions, even in their outskirts. The 3.3 keV line is thought to be a combination of  {\sc Ar XVIII}, {\sc S XVI} and of K~K$\alpha$ instrumental lines, and there are two weaker lines at 3.7~keV, potentially an {\sc Ar XVII} complex and/or instrumental Ca K$\alpha$ line. Both the 3.3 and 3.7 keV lines have been detected in other deep {\it XMM-Newton} observations, several years ago (c.f.~\cite{Boyarsky:14a,Boyarsky:2014ska,Ruchayskiy:2015onc}). When including the 3.3 and 3.7 keV lines, the SM in DRS finds that the limit on DM flux is relaxed by a factor of $\sim$8. 

To test the potential presence of the 3.3 and 3.7 keV lines in the DRS data, I fit to their published spectra for both the PN and MOS observations including these lines and a power-law background. For the PN data, the 3.3 and 3.7 keV lines were preferred at 4.3$\sigma$ and 2.8$\sigma$, respectively. For the MOS data, the 3.3 and 3.7 keV lines were preferred at 3.5$\sigma$ and 1.6$\sigma$, respectively. When including both lines, the bounds on the inclusion of the 3.5 keV line were relaxed at about a factor of $\sim$10 for both the PN and MOS observations, and this is consistent with the SM results in DRS, which used the full framework of their joint-likelihood analysis. Shown in Fig.~1 are example a cases where the 3.5 keV is at its 95\% upper limit flux for both the PN and MOS instruments, which shows how the data prefers the presence of the 3.3 and 3.7 keV lines, and could include the 3.5 keV line with appreciable flux. 

\begin{figure}
    \centering
    \includegraphics[width=6.5in]{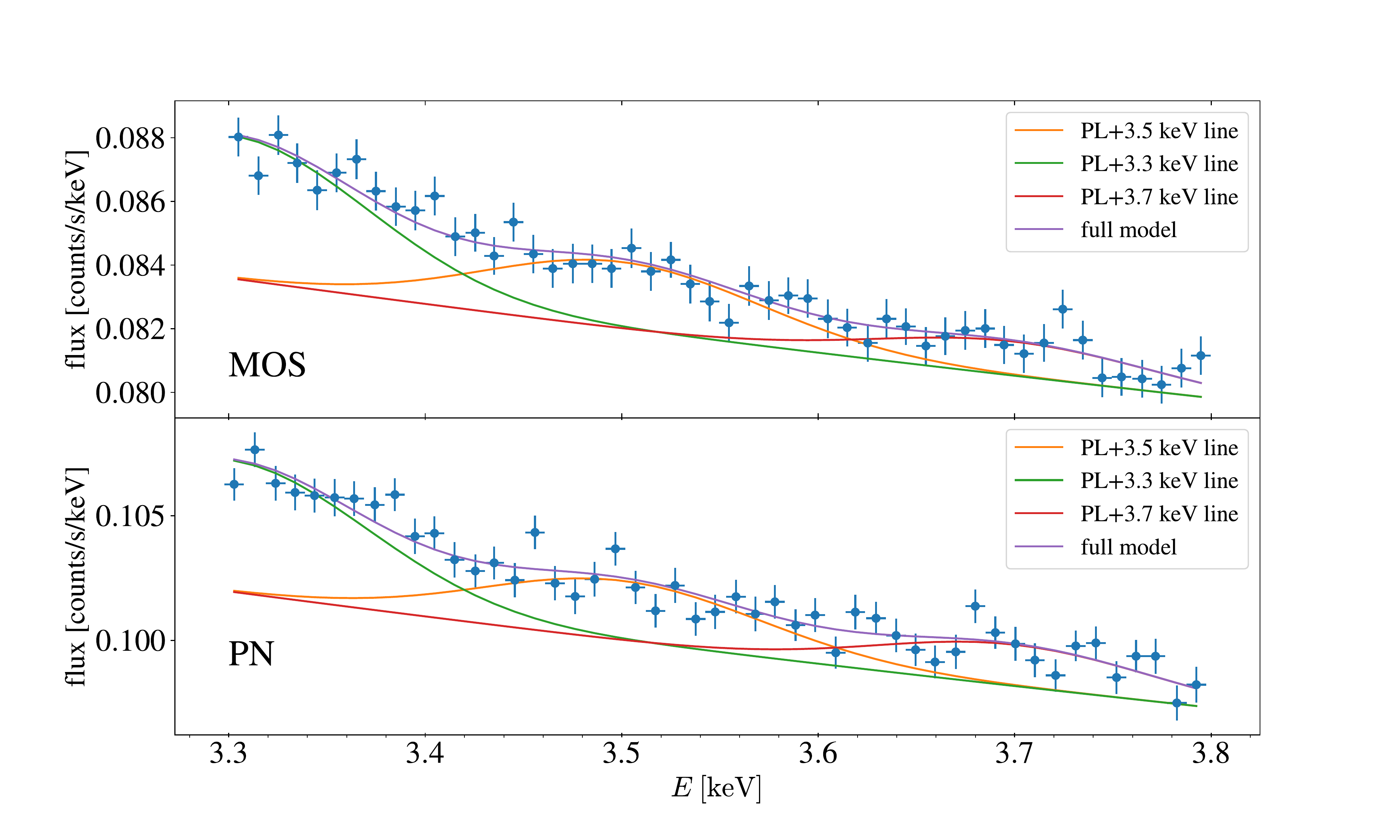}
    \caption{The data from DRS are shown in blue points with errors. I refit to this data and find preference for the extra 3.3 and 3.7 keV lines at 4.3(3.5)$\sigma$ and 2.8(1.6)$\sigma$, for the PN (MOS) case, respectively. In the models shown here, I take the 3.5 keV line to be at its 95\% limit, with the corresponding best fit power law and 3.3 \& 3.7 keV lines. Model components are shown with the power law (PL) plus each line individually, plus the full model.}
    \label{fig:spectra}
\end{figure}

The preference for the inclusion of the 3.3 and 3.7 keV lines begs the question as to what the results would be when including data outside of the 3.3-3.8 keV window, which would further constrain the presence of these lines as well as others at 3.1 keV, 3.9 keV and 4.1 keV, and provide for a more complete measure of the potential presence of the 3.5 keV line. A broader-spectrum analysis is certainly called-for given the presence of lines within the energy region of interest, but is beyond the scope of this comment. Such broader analysis could provide stronger constraints, or potentially detect the 3.5 keV line as other work with archival {\it XMM-Newton} data has done \cite{Boyarsky:2018ktr}.

\begin{figure}
    \centering
    \includegraphics[width=6.5in]{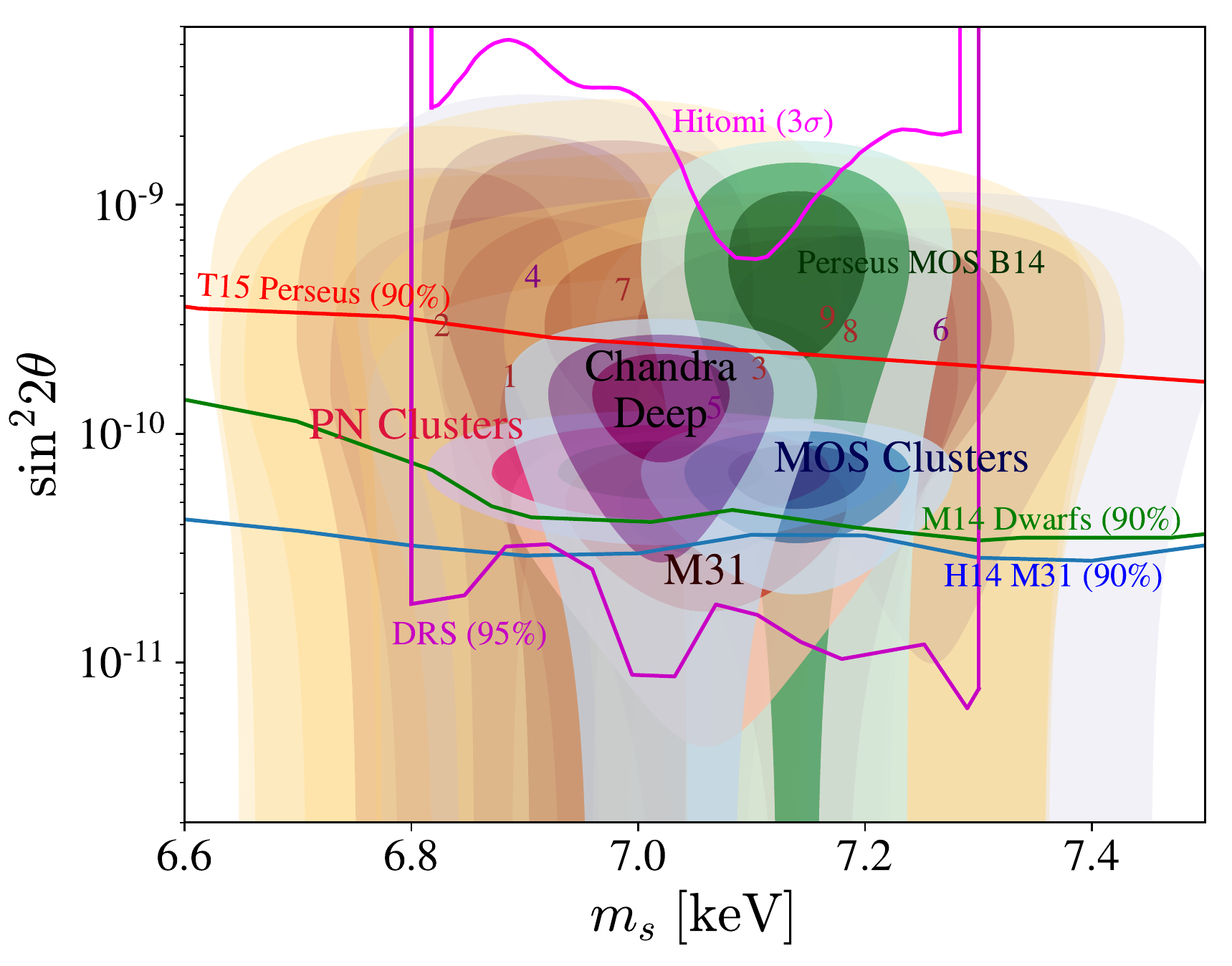}
    \caption{Shown here is the newly derived limit inferred from DRS when including the DM density profile uncertainties and the presence of lines at 3.3 and 3.7 keV. I use the extra lines limit from the SM of DRS, combined with the DM density profile uncertainties presented here. Also shown are the parameters consistent with previous observations' detections of the 3.5 keV line as well as commensurate limits (for description, see Abazajian (2017), \cite{Abazajian:2017tcc}). Expanding the narrow energy window studied in DSR may enhance or relax the constraint further.}
    \label{fig:parameters}
\end{figure}

\section*{Conclusion}
I have highlighted a number of model choices in the DM density profile and X-ray spectral line backgrounds that relax the constraints claimed in DRS by at least a factor of $\sim$20. Many of these model choices, but not all, were included in the SM of DRS, though the conclusions of DRS did not commensurately reflect the model dependency of their claimed limit. Appropriate conservative model selection relaxes the claimed constraints to at least those shown in Fig.~2. When expanding the spectral region to further characterize all components as well as the potential DM contribution, the constraint may relax even further, could strengthen, or even reveal a 3.5 keV line consistent with other observations.



\bibliography{scibib}

\begin{thebibliography}{10}

\bibitem{Abazajian:2017tcc}
K.~N. Abazajian, {\it Phys.\ Rept.\/} {\bf 711-712}, 1 (2017).

\bibitem{Dessert:2018qih}
C.~Dessert, N.~L. Rodd, B.~R. Safdi, {\it Science\/} {\bf 367}, 1465 (2020).

\bibitem{Zhang:2012rsb}
L.~Zhang, {\it et~al.\/}, {\it Astrophys. J.\/} {\bf 772}, 108 (2013).

\bibitem{Kuhlen:2012qw}
M.~Kuhlen, J.~Guedes, A.~Pillepich, P.~Madau, L.~Mayer, {\it Astrophys. J.\/}
  {\bf 765}, 10 (2013).

\bibitem{Weinberg:2001gm}
M.~D. Weinberg, N.~Katz, {\it Astrophys. J.\/} {\bf 580}, 627 (2002).

\bibitem{Robles:2019mfq}
V.~H. Robles, T.~Kelley, J.~S. Bullock, M.~Kaplinghat, {\it Mon.\ Not.\ Roy.\
  Astron.\ Soc.\/} {\bf 490}, 2117 (2019).

\bibitem{Boyarsky:14a}
A.~Boyarsky, O.~Ruchayskiy, D.~Iakubovskyi, J.~Franse, {\it Phys.Rev.Lett.\/}
  {\bf 113}, 251301 (2014).

\bibitem{Boyarsky:2014ska}
A.~Boyarsky, J.~Franse, D.~Iakubovskyi, O.~Ruchayskiy, {\it Phys.\ Rev.\
  Lett.\/} {\bf 115}, 161301 (2015).

\bibitem{Ruchayskiy:2015onc}
O.~Ruchayskiy, {\it et~al.\/}, {\it Mon.\ Not.\ Roy.\ Astron.\ Soc.\/} {\bf
  460}, 1390 (2016).

\bibitem{Boyarsky:2018ktr}
A.~Boyarsky, D.~Iakubovskyi, O.~Ruchayskiy, D.~Savchenko  (2018).
  ArXiv:1812.10488.

\end{thebibliography}

\bibliographystyle{Science}

\section*{Acknowledgments}
I acknowledge useful discussions with Ben Safdi, who agrees that the conservative interpretation of the limits from DRS are approximately consistent with those in Fig.~2. I also acknowledge helpful comments on this manuscript from Alexey Boyarksy, Esra Bulbul, George Fuller, Manoj Kaplinghat, Alex Kusenko, and Oleg Ruchayskiy, as well as support from NSF Theoretical Physics Grant PHY-1915005.

\end{document}